\newcommand{\un}{~\mathrm}

\documentclass[twocolumn
,prl
,floatfix
,showpacs
,preprintnumbers
,amsmath
,amssymb
]{revtex4}

\usepackage{graphicx}
\usepackage{dcolumn}
\usepackage{bm}
\usepackage{times}
\usepackage{natbib}
\usepackage{color}
\definecolor{Blue}{rgb}{0.3,0.3,0.9}
\definecolor{Red}{rgb}{0.9,0.3,0.3}
\definecolor{Green}{rgb}{0.3,0.9,0.3}

\begin{document}

\title{Toughening and asymmetry in peeling of heterogeneous adhesives}
\author{S. Xia}
\affiliation{Woodruff School of Mechanical Engineering, Georgia Institute of Technology, Atlanta, GA 30332-0405, USA}
\author{ L. Ponson}\email{laurent.ponson@upmc.fr}
\affiliation{CNRS, UMR 7190, Institut Jean Le Rond d'Alembert, F-75005 Paris, France}
\affiliation{UPMC Univiversit\'e Paris 06, UMR 7190, Institut Jean Le Rond d'Alembert, F-75005 Paris, France}
\author{G. Ravichandran and K. Bhattacharya}
\affiliation{Division of Engineering and Applied Science, California Institute of Technology, Pasadena CA 91125, USA} 

\begin{abstract}
The effective adhesive properties of heterogeneous thin films are characterized through a combined experimental and theoretical investigation. By bridging scales, we show how variations of elastic or adhesive properties at the microscale can significantly affect the effective peeling behavior of the adhesive at the macroscale.  Our study reveals three elementary mechanisms in heterogeneous systems involving front propagation: (i) patterning the elastic bending stiffness of the film produces fluctuations of the driving force resulting in dramatically enhanced resistance to peeling; (ii) optimized arrangements of pinning sites with large adhesion energy are shown to control the effective system resistance, allowing the design of highly anisotropic and asymmetric adhesives; (iii) heterogeneities of both types result in front motion instabilities producing sudden energy releases that increase the overall adhesion energy. These findings open potentially new avenues for the design of thin films with improved adhesion properties, and motivate new investigation of other phenomena involving front propagation.
\end{abstract}

\date{\today}
\maketitle
Bridging microscale properties of materials with their effective mechanical behavior at the macroscale is a major challenge in both pure and applied science. A great deal of research effort has been dedicated to the study of effective elastic properties in heterogeneous systems. Composite materials, structures and meta-materials can achieve extraordinary effective properties, {\it e. g.} negative Poisson's ratio \cite{Rothenburg} or stiffness greater than diamond \cite{Jaglinski}, and elegant homogenization techniques have been developed in order to efficiently link micro to macroscale in elastic settings \cite{Milton,NematNasser,Sanchez}. Surprisingly, our understanding of the role of heterogeneities on the overall resistance of these systems to failure resulting from the propagation of free boundaries and free discontinuities is rather limited, despite the major importance of this question in engineering science. For example, stress concentration generated by brittle cracks makes the macroscopic system extremely sensitive to microscopic features - the scale invariant roughening of cracks as well as their highly intermittent dynamics are good illustrations of this effect \cite{Ponson5,Bonamy5} - raising fundamental impediments to the development of reliable homogenization techniques for fracture problems. 

In this letter, we address the challenge of finding the macroscopic resistance to front propagation in a heterogeneous media in the context of thin film peeling. There has been a recent renewal of interest in adhesive systems driven by the exploration of exceptional properties of biological systems like geckos.   Consequently, much attention has focused on adhesion enhancement achieved by 3D features of the adhesive surface, such as arrays of fibrils \cite{Autumn,Geim} or hierarchical structures \cite{Yao}. Here, we will show that such a rather complex microstructure is not required to achieve enhanced adhesion: heterogeneities in the elastic properties of the film and variations of the frature energy introduced on the flat adhesive surface can lead to similar, and even better, performance.

\begin{figure}
\includegraphics[width=1\columnwidth]{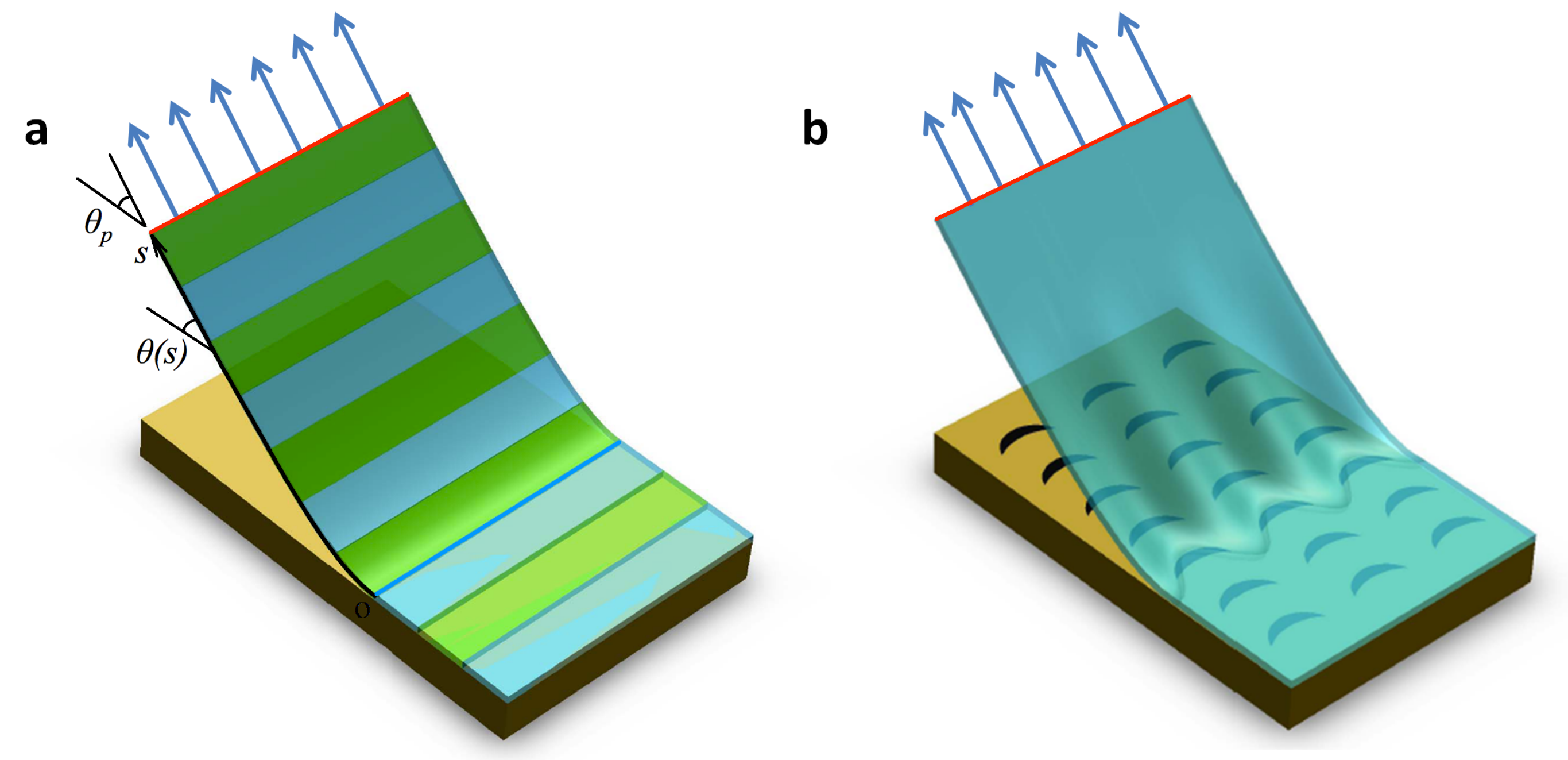}
\centering
\caption{Peeling of adhesive tapes with (a) elastic and (b) adhesive heterogeneities.}
\label{Fig1}
\end{figure}

Taking advantage of the rather simple geometry of peeling thin films, we have designed an experimental set-up (see Fig.~\ref{Fig1}) for which the microscale material properties are fully controlled and tunable. 
First, we consider thin films with inhomogeneous elastic properties (see Fig.~\ref{Fig1}(a)), and show that these  inhomogeneous elastic properties can be used to control both the effective peeling force and the overall adhesion energy without any modification of the adhesive surface of the film. Second, we investigate systems with heterogeneous adhesive properties (see Fig.~\ref{Fig1}(b)), and show that these can be used to produce strength anisotropy and asymmetry. We then discuss optimizing the microstructure for strong macroscopic effects. We conclude with a discussion of applications of these concepts to other front propagation systems.

\begin{figure}
\includegraphics[width=.9\columnwidth]{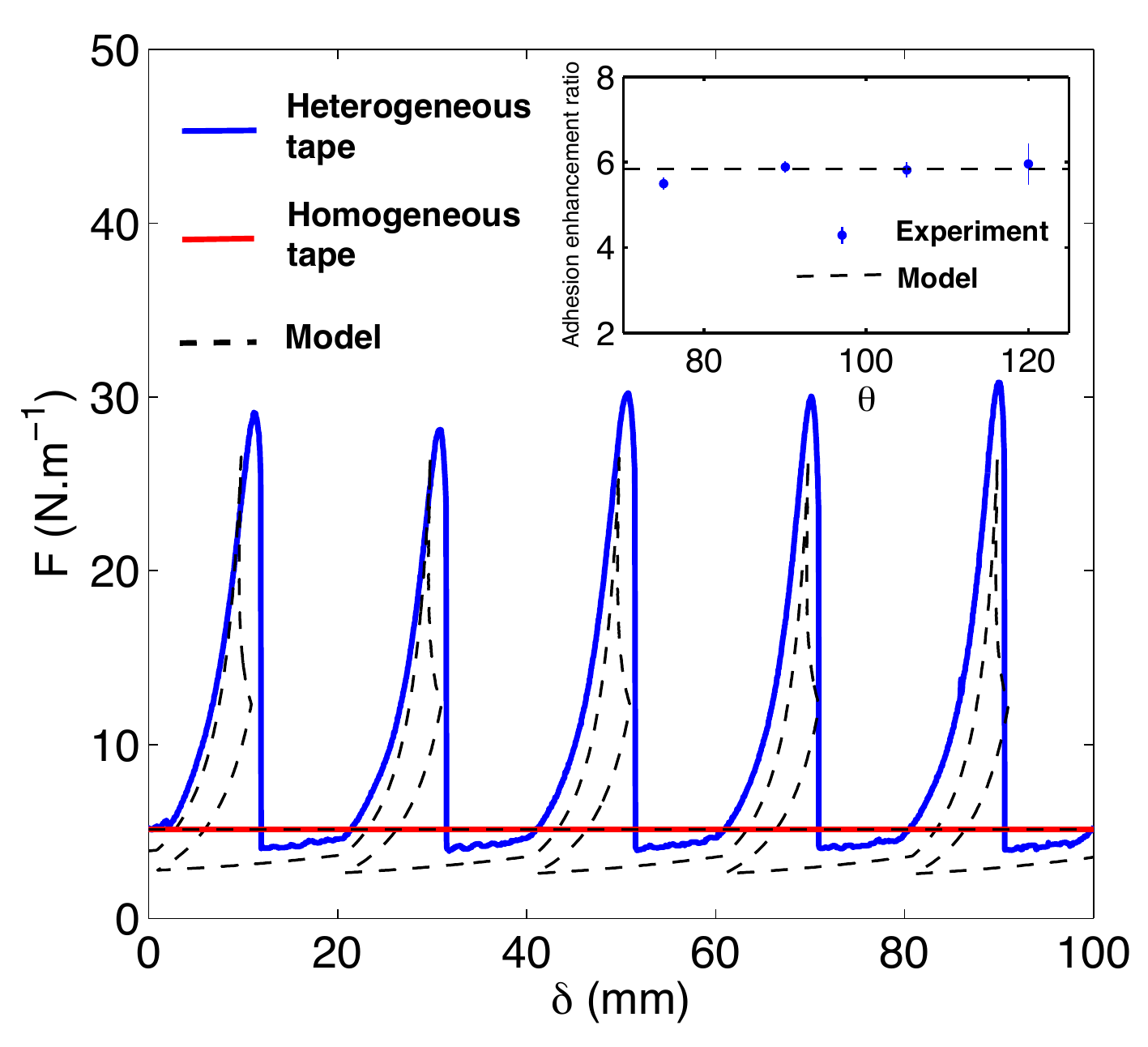}
\centering
\caption{Measured force required to peel a tape of alternating bending stiffness (blue solid line) compared to that required to peel a homogenous tape (red solid line) as well as theoretical predictions for both materials (black dashed lines). Here, we impose $\theta_p = 90^\circ$ for peel angle at the extremity of the tape. The inset shows the variations of the enhancement factor $\frac{F_{\mathrm{het}}}{F_{\mathrm{hom}}}$ as a function of the imposed peel angle for experiments and model.}
\label{Fig2}
\end{figure}

In the first set of experiments, we consider thin films with alternating stiff and compliant stripes peeled from a rigid substrate with uniform adhesion ($G_c = 5.1 \un{J.m^{-2}}$). This heterogeneous tape is made by gluing patches of thin polyester film of thickness $\Delta h = 130\un{\mu m}$ on the back-side of a continuous polyester film of thickness $h_0 = 161 \un{\mu m}$ \cite{Footnote}.  As reference, we use two thin films of constant thickness,  one with uniform thickness $h_0$ and one with uniform thickness $(h_0+\Delta h)$.  We peel these tapes from an epoxy substrate that can be considered to be perfectly rigid relative to the film compliance.  The extremity of the film is pulled at a constant velocity $v_m$ (indicated by arrows in Fig.~\ref{Fig1}(a)) while maintaining a constant peel angle $\theta_p$ between the mean plane of the film and the horizontal plane. This causes the film to peel, and the force required to maintain the constant velocity is measured and shown in Fig.~\ref{Fig2} (the peel force is normalized by the width of the tape in the figure).   The force required to peel the two reference homogeneous tapes coincide as anticipated by Rivlin \cite{Rivlin}.  However, the force required to peel the heterogeneous tape oscillates as the peel front negotiates the stiff and compliant regions, with the peaks attained as the peeling front goes from a compliant to a stiff portion of the tape.  The effective force required to peel a macroscopic length of this tape is equal to the maximum measured force.  Indeed, the peeling front would get stuck if the applied force were smaller than the peak force. Therefore, we refer to the peak force as the {\it effective peel force}.  Remarkably, the effective peel force is $29.0 \un{N.m^{-1}}$ which is almost six times the force of $5.1 \un{N.m^{-1}}$ required to peel either homogenous tape.  The experiment is repeated at various imposed peeling angles $\theta_p$, and no effect is noticeable in the enhancement factor (see inset of Fig.~\ref{Fig2}).

To understand why patterning the elastic stiffness of the tape significantly enhances its stickiness, we treat the film as inextensible and apply Griffith's criterion \cite{Griffith}.  Accordingly, the driving force per unit width for peeling is given by the incremental work $\mathrm{d}W$ of the external loading system and the variations of the stored elastic energy $\mathrm{d}U$ (due to bending):  $G = \frac{1}{b} \frac{d(W - U)}{dc}$ where $c$ is the peel front position. The peel front propagates when this driving force equal a critical value, $G=G_c$.  For homogeneous thin films, only the first contribution is relevant to the driving force since $U$ remains unchanged as peeling progresses.  This so-called Rivlin model \cite{Rivlin} leads to the expression of the peeling force per unit width
\begin{equation}
F_{\mathrm{hom}} = \frac{G_\mathrm{c}}{1-\cos \theta_p}.
\label{Eq1}
\end{equation}
Note that this behavior is independent of the elastic properties of the tape as we observe in Fig.~\ref{Fig2} for the homogenous systems.

On the contrary, for heterogenous elastic properties, the stored elastic energy varies rapidly as the peeling front crosses from a compliant to a stiff region. To describe this effect quantitatively, the tape is modeled by an unextensible heterogeneous Euler-Bernoulli beam, and its geometry as peeling proceeds is described by the angle $\theta (s)$ that the tangent to the film at a distance $s$ from the peeling front makes with the horizontal plane of the epoxy substrate (see Fig. 1(a)). The latter is governed by the equation \cite{Antman,Landeau}
\begin{equation}
 \frac{d}{ds} \left ( D(s) \frac{d \theta}{ds}  \right ) = b F \sin(\theta_p - \theta(s) )   
\label{Eq1a}
\end{equation}
where $D(s)$ is the bending rigidity of the film at position $s$, $F$ is the applied peeling force per unit width and $\theta_p$ is the imposed peel angle at the extremity of the film. We solve this equation semi-analytically, and the computed variation of peeling force is compared with the experimental data in Fig.~\ref{Fig2}. The effective or peak force as the peel front crosses over from the compliant to the stiff region can be calculated exactly for a film with large periodicity $\lambda$ of the heterogeneities with respect to characteristic material length $r_b = \sqrt{\frac{D_c}{2 b G_c}}$ (the radius of curvature of a tape of a homogenous compliant material during peeling). It follows
\begin{equation}
\frac{F_{\mathrm{het}}}{F_{\mathrm{hom}}} = \frac{D_{\mathrm{s}}}{D_\mathrm{{c}}}
\label{Eq2}
\end{equation}
where $D_{\mathrm{s}}$ (resp. $D_{\mathrm{c}}$) is the bending rigidity of the stiff (resp. compliant) region.  The enhancement ratio is predicted to be independent of peel angle in agreement with experimental observation (see inset of Fig.~\ref{Fig2}).  Further, the enhancement ratio depends on the ratio of the bending rigidity $D = \frac{E b h^3}{12(1-\nu^2)}$ which in turn depends on the third power of the thickness, $h$. This is the reason why modulating the thickness of the tape has a significant effect on the adhesion.

The mechanism of peeling force enhancement is clear from both the experiment and the model: as the peeling front traverses from the compliant to the stiff material, a significant portion of the work done by the peel force goes into bending, the suddenly stiffer region draining energy away from the peeling front. This gives rise to a drop in the total driving force for peel resulting in a peak in the peel force when a constant velocity is imposed. Similarly, the dip as the front traverses from the stiff to the compliant region is explained by the sudden release of bending energy. Bending energy is small, but its rate of change is large enough to create dramatic effects.

\begin{figure}
\includegraphics[width=1.\columnwidth]{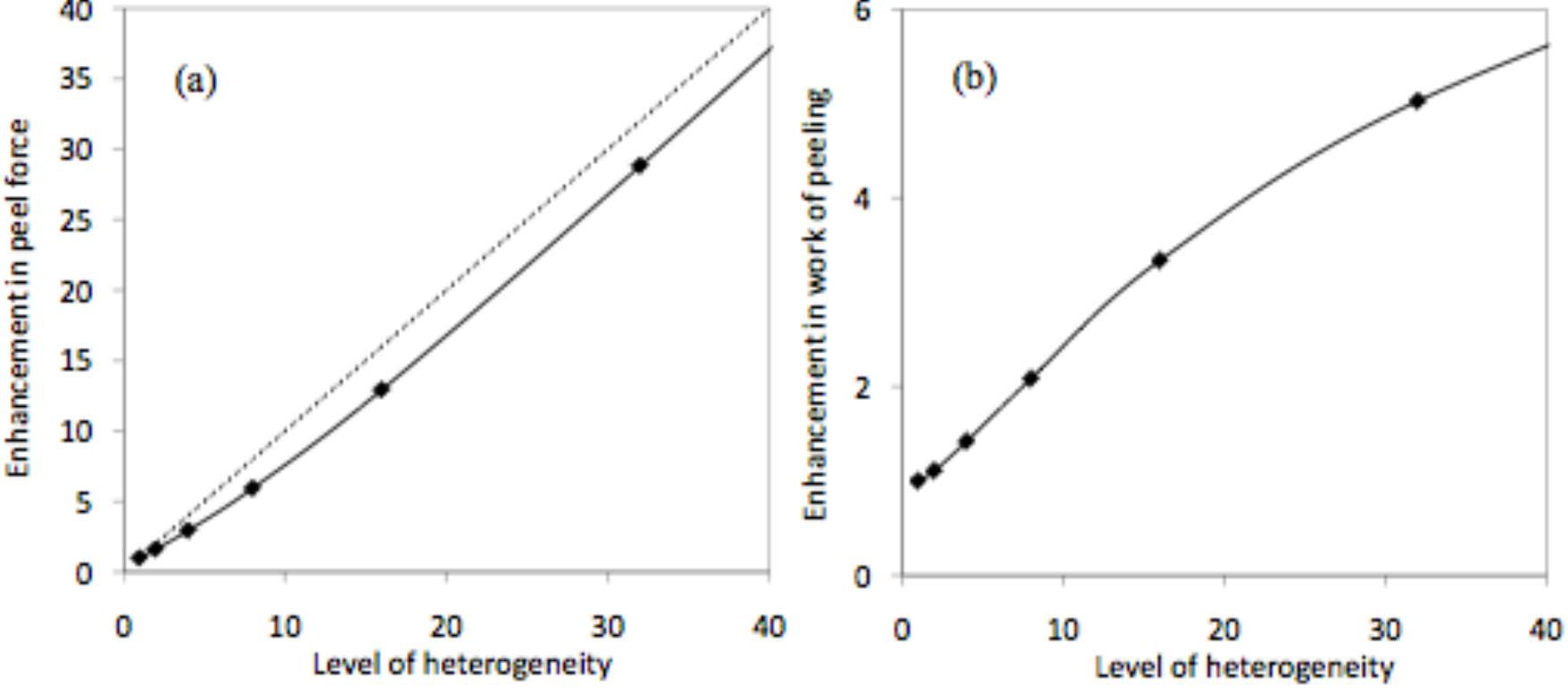}
\centering
\caption{Toughening in elastically heterogeneous adhesives: (a) Predicted enhancement factor $\frac{F_{\mathrm{het}}}{F_{\mathrm{hom}}}$ achieved on the macroscopic peeling force as a function of the level of heterogeneity $\frac{D_s}{D_c}$. The analytical prediction of Eq. (\ref{Eq2}) is represented by the dotted line for the large period case ($\lambda \gg r_b$), and the numerical solution for a small period of heterogeneity is shown by the solid line ($\lambda \ll r_b$). (b)  Predicted enhancement factor $\frac{G_\mathrm{c}^{\mathrm{eff}}}{G_\mathrm{c}}$ achieved on the effective adhesion energy as a function of the level of heterogeneity $\frac{D_s}{D_c}$.}
\label{Fig3}
\end{figure}

Fig.~\ref{Fig3}(a) shows the enhancement ratio for various levels of heterogeneity. The analytic prediction of Eq. (3) for large periode ($\lambda \gg r_b$) is compared with that of a film with a small period ($\lambda \ll r_b$). The latter is obtained by numerically solving Eq. (2). Note that the enhancement drops for a film with small period because the region being bent spans a number of periods and the peeling zone senses the effective bending stiffness which is strictly larger that the stiffness of the compliant region.

We look now at the mechanism of peeling from an energy perspective, and investigate the effective adhesion energy $G_\mathrm{c}^{\mathrm{eff}}$ required to peel the film over a macroscopic length. For a quasi-static process, this energy is the average work $\frac{1-\cos \theta_p}{\lambda b} \int_0^{\lambda} F(\delta) d\delta$ of the external force required to peel the front over one period $\lambda$. It is given by the area below the curve $F(\delta)$ of Fig.~\ref{Fig2}. Obviously, the effective adhesion energy as measured in the experiments is larger for the heterogeneous tape than that for the homogeneous one. 

To understand the mechanisms leading to this toughening, we need to look more closely at the theoretical curve $F(\delta)$ for heterogeneous tapes shown in Fig.~\ref{Fig2}.  Note that $F(\delta)$ is multivalued:  this is because the theoretical calculation is indexed by the position of the peel front along the tape.  However, in the experiment, the displacement is imposed and increases monotonically.  Consequently the snap-back branches are not observed; instead there is an abrupt drop of the force accompanied by a sudden jump of the peeling front from one position to another. Therefore, the energy that is stored in the bending region while the peel force increases is not recovered (this recovery corresponds in the theoretical model to the descending branch), but instead partly dissipated as heat, acoustic emissions etc.  We compute the area under the curve and describe this as the {\it effective adhesion energy} $G_\mathrm{c}^{\mathrm{eff}}$.   The predicted toughening factor $\frac{G_\mathrm{c}^{\mathrm{eff}}}{G_\mathrm{c}}$ is shown in Fig.~\ref{Fig3}(b) as a function of the level of heterogeneity $\frac{D_s}{D_c}$.  Note that this represents toughening under quasistatic conditions.  Under dynamic conditions, the drops would not be abrupt and the amount of energy dissipated would be larger.

We emphasize that the significant enhancement of the effective force and the effective adhesion energy do not result from improvements made at the adhesive interface, but through variations of the structural properties of the film.  We remark that the toughening mechanism and instabilities are generic and not limited to peeling of thin films. The peeling of a stiff plate from a compliant substrate can lead to oscillations in the peeling moment in the presence of heterogeneities \cite{Ghatak2}. Heterogeneities of fracture energy in 3D brittle materials have been shown to lead to similar instabilities, producing the so-called crackling noise \cite{Sethna,Maloy2,Bonamy5}. This effect was shown to increase considerably the macroscopic fracture energy of brittle disordered materials \cite{Roux5,Ponson14}. 
\vspace{\baselineskip}

We now turn to the second experiment where we keep the elastic modulus of the film uniform but pattern the adhesive energy in order to investigate the effect of microscale heterogeneities at the film-substrate interface on the overall peeling behavior. We mold a polydimethylsiloxane (PDMS) film on to a rigidly supported polyester transparency sheet patterned with a periodic array of ink features using a regular laser printer. This produces a well-controlled heterogeneous tape as shown schematically in Fig.~\ref{Fig1}(b), since the adhesive energy of the region with the printed ink is about four times larger than that of the bare transparency sheet (3.55 vs. $0.65\un{J.m^2}$ at a peel front velocity of $1.0\un{\mathrm{\mu}m.s^{-1}}$).). As we peel the film, the peeling front becomes wavy as it tries to go ahead in the weaker region as shown in Fig.~\ref{Fig1} (b). This results in additional bending of the debonded part of the film (see the corrugations of the tape in the figure), the elastic energy of which balances the gain in fracture energy.

We model this problem analytically by treating the film as a (finite deformation) Kirchhoff plate where the film is assumed to be inextensible in the plane and the stored energy density is proportional to the curvature. If the peeling front were straight, then the film being peeled would bend only in one direction, and Kirchhoff plate theory would reduce to the Euler-Bernoulli beam modified by Poisson's ratio. We assume that the peeling front displays small geometrical perturbations $f(x,t)$ with respect to the straight configuration ($\vec{x}$ is here parallel to the peeling front), so that the shape of the film is close to the singly curved solution associated with the straight front. We linearize the Kirchhoff plate theory around the singly curved solution to obtain the shape of the film, and then calculate the total deformation energy of the film. The perturbation $\delta G(x)$ of the force driving the peeling front is obtained by calculating the variation of this energy with respect to an infinitesimal increment of the peel front. We obtain an evolution law for the peeling front by postulating that the normal velocity is proportional to the net driving force $G_0 + \delta G - G_\mathrm{c}$ in which $G_0$ is the driving force of the straight front. It is convenient to express this in terms of the Fourier transform with respect to $\vec{x}$ (hat represents the transform and $k$ is the Fourier variable).  We obtain
\begin{equation}
\frac{\partial \hat f}{\partial t} \propto g(k) \hat f  + \hat G_\mathrm{f}
\label{Eq5}
\end{equation}
where $g(k) = - \alpha_0 |k|$ except for very small $k$ with $\alpha_0 = 2 F \tan^2 \theta_p $, and
$G_\mathrm{f} = F ( 1 - \cos \theta_p ) - G_\mathrm{c}(x,f(x,t),\dot{f}(x,t))$. In our experiments, $G_\mathrm{c}$ is slightly rate dependent and we use $G_\mathrm{c} \propto (\dot{f}) ^n$ with $n\simeq 0.3$ on the basis of measurements on homogeneous ink or polyester interfaces. We can obtain a relation between average velocity and the overall force by averaging over $x$, or equivalently looking at the limit $k\to 0$.  We obtain, $F \simeq \frac{ \langle G_\mathrm{c}(x,f(x,t),\dot{f}(x,t)) \rangle_\mathrm{x} } {1-\cos \theta_p} $. Consequently, the shape $f(x,t)$ and the velocity $\dot{f}(x,t)$ of the front that are collective responses to the overall pattern govern the overall peeling force. This opens the door for the design of adhesives with heterogeneities patterned to achieve desired effective peeling blue behavior.

We exploit this idea to create asymmetry. It has been recognized that one can introduce anisotropy whereby fronts propagating in different directions have different adhesion strengths \cite{Ramrus,ChenB}. However, it is possible to go further, and show that the effective adhesive strength depends not only on the direction but also on the sense. In other words, the adhesive strength of peeling a tape from left to right can be significantly different than that of peeling from right to left.   This is accomplished by creating pinning sites with asymmetric geometry so that the shape of the front as it navigates the pinning sites in one direction is different from that as it goes in the opposite direction. The effective force is different since it depend on the shape of the front.

\begin{figure}
\includegraphics[width=1.\columnwidth]{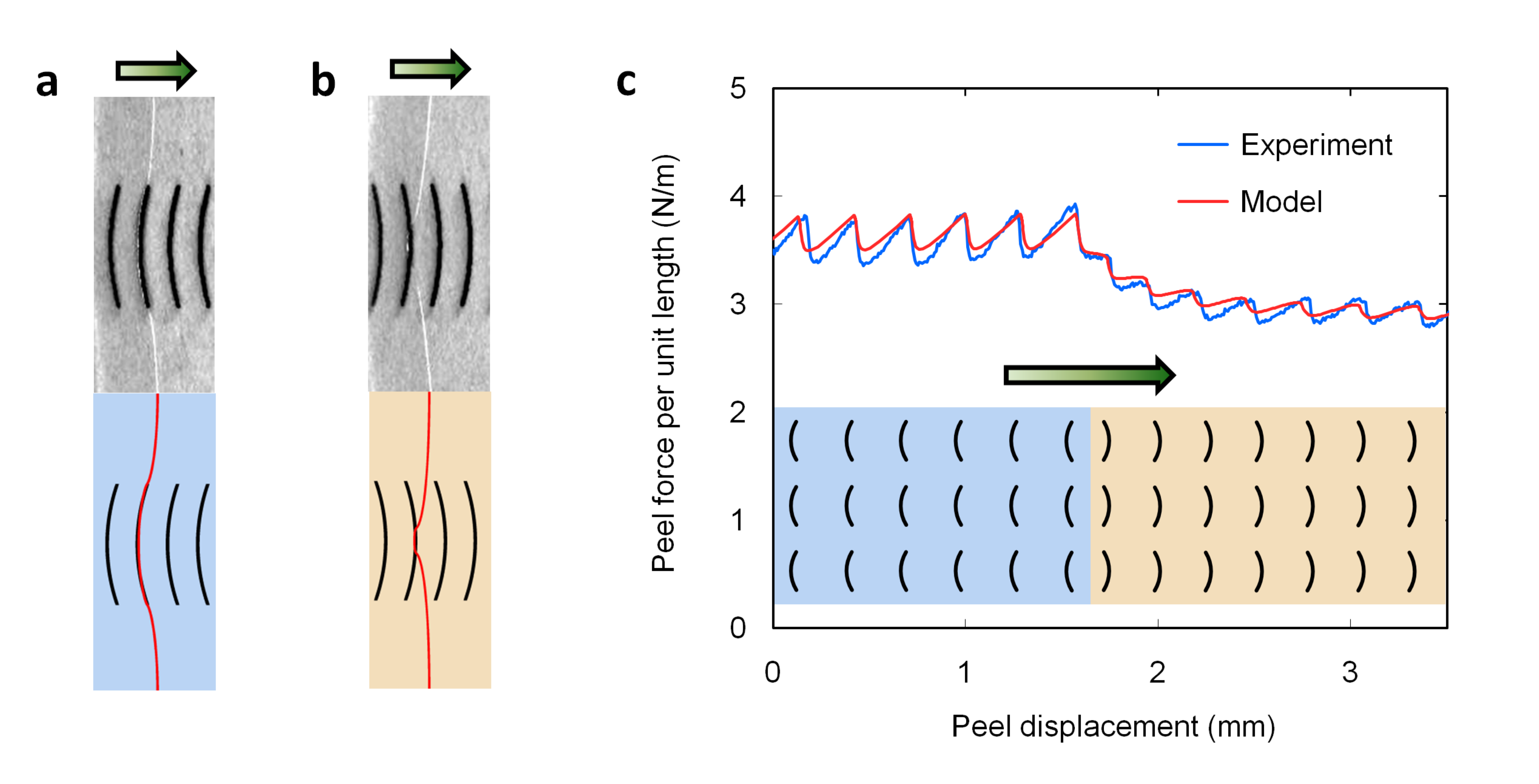}
\centering
\caption{Exceptional asymmetry due to adhesive strength heterogeneity: (a,b) The peeling front is distorted as it negotiates regions of enhanced/diminished adhesion energy. Top halves show the experimental observations while the bottom halves show the results of the theory. The arc features have a constant line thickness of $100 \un{\mu m}$ and a vertical span of $4 \un{mm}$. (c) The measured force to peel a film from a patterned substrate: note that the force in the forward-facing region is higher than that in the backward-facing region.}
\label{Fig4}
\end{figure}

We demonstrate this idea with the arc pattern shown in Fig.~\ref{Fig4}. We create two regions with the pattern pointing in opposite directions. This allows us to examine the adhesive strength in both directions in a single test and removes artifacts resulting from variations in sample preparation and loading. As we peel the film, the peel force oscillates as it passes each column of arc-shaped regions of higher adhesion energy. More importantly, the effective adhesive strength in the forward region (front first touching the convex portion of the arc) is $25 \un{\%}$ higher than that in the backward region. As the front propagates in the forward region, it first encounters the curved convex portion of the arc and sees them as significant obstacles that require a large peeling force to overcome.
On the other hand, as the front propagates in the backward region, it first encounters the narrow arms of the arc and sees these as smaller obstacles.  The evolution equation Eq. (\ref{Eq5}) is used to model this phenomenon. The computed front shape as well as the peel forces are also shown in Fig.~\ref{Fig4} and agree with the experimental observations. 

In order to achieve stronger asymmetry, we can optimize the pattern. Using an algorithm developed in the context of elastic manifolds driven in random media \cite{Rosso}, the strength of asymmetry is computed efficiently allowing the exploration of a large range of geometrical parameters and strength contrast. It shows that heterogeneous tapes with resistance at least twice larger in one direction than the opposite one can be designed.


The significance of these results to adhesion is clear: exceptional strength, anisotropy and asymmetry. 
But it extends beyond, since we may regard adhesion fronts as a prototypical problem in condensed matter physics.
The equation of motion for the peeling front (Eq. (\ref{Eq5})) arises in the study of myriad phenomena including brittle fracture \cite{Rice4,Leblond}, dislocations \cite{Hirth}, phase boundaries \cite{Dondl} and wetting fronts \cite{Joanny}. 
Most of the research effort has focused on the disordered case where $G_\mathrm{c}$ is described by a quenched noise, resulting in universal features through intermittent dynamics and scale-invariant roughening \cite{Bonamy6,Sethna}. 
Our work shows that there is much to gain in terms of overall properties by studying the deterministic and periodic cases.  Specifically it identifies three distinct mechanisms: (i) patterning the elastic bending stiffness produces fluctuations of the driving force resulting in largely enhanced resistance to peeling; (ii) optimized arrangements of pinning sites with large adhesion energy are shown to control the effective system resistance allowing significant anisotropy and asymmetry; (iii) heterogeneities of both types result in front instabilities producing sudden energy release that increase the overall dissipated energy in the system.  Together these can potentially open the door to engineering new materials where the toughness, strength, etc., can be tuned through designed defects \cite{Xia}.

\paragraph{Acknowledgement}  This work was conducted when SX and LP held post-doctoral positions at Caltech. We thank J.-B. Leblond, P. Mithal and A. Rosso for helpful discussions. The financial support from the US National Science Foundation (all authors) and from the European Union through the PhyCrack Marie Curie Fellowship (LP) is gratefully ackowledged.

\bibliographystyle{apsrev}


\begin{thebibliography}{23}
\bibitem{Footnote}
Both the patches and the base films have width $b=38.0 \un{mm}$, Young's modulus $E = 3.8 \un{GPa}$, Poisson's ratio $\nu = 0.42$ and the spacing of the patches is  $\lambda = 20 \, \un{mm}$.
\end{thebibliography}

\end{document}